\documentclass[sigconf,screen] {acmart}
\setcopyright{none}
\copyrightyear{2026}
\acmYear{2026}
\acmConference[Workshop 'Human-Centered AI for Expressive Arts Therapy' at DIS Companion '26]{Designing Interactive Systems Conference}{June 13--17, 2026}{Singapore, Singapore}
\setcopyright{none}
\usepackage[utf8]{inputenc}
\usepackage{graphicx}

\usepackage{subcaption}
\usepackage{enumitem}
\usepackage{array,booktabs}
\acmISBN{}
\acmDOI{}
\newcolumntype{L}[1]{>{\raggedright\arraybackslash}p{#1}}
\newcolumntype{Y}{>{\raggedright\arraybackslash}X}

\AtBeginDocument{}
\renewcommand\footnotetextcopyrightpermission[1]{}

\begin{document}

\title{Measuring the "Interaction Gap" in Drama Therapy with AI}

% 저자 정보
\author{Sora Kang}
\email{sorakang@snu.ac.kr}
\affiliation{%
  \institution{Seoul National University}
  \city{Seoul}
  \country{Republic of Korea}
}

% --- CCS Concepts ---


% --- Keywords ---

% 단축 저자명
\renewcommand{\shortauthors}{Sora Kang}

\begin{abstract}
Generative AI is increasingly being introduced into expressive arts therapy, where it is often credited with offering a non-judgmental environment that supports psychological safety. Existing HCI work has largely positioned AI as a co-creative material or as a bridge/mediator into human-led care. This paper explores a different position. When a patient performs the same drama therapy task with an AI partner and with a human partner, the resulting self-presentations tend to differ in patterned ways. We propose treating this difference, the Interaction Gap, as a diagnostic lens within drama therapy. Rather than asking which context elicits a truer self, the lens reads the difference between the two performances as information about the social pressures shaping self-expression in each context. We sketch a starting point for task design and measurement signals grounded in drama therapy's existing use of role and aesthetic distance, and raise provocations for workshop discussion: the observer effect and privacy paradox that measurement introduces, and the question of whose lens the gap is.
\end{abstract}

\maketitle

\section{Introduction}
Expressive arts therapy uses visual art, music, dance, and drama to support emotional awareness, expression, and self-understanding~\cite{malchiodi2005expressive, jones2007drama}. Recent generative AI systems, capable of producing images, music, text, and soundscapes across modalities, have entered these practices as a new medium and collaborator. A growing body of HCI work reports that AI's perceived non-judgmental quality facilitates self-disclosure, with users describing AI partners as offering a sense of emotional sanctuary~\cite{siddals2024happened, lucas2014itsonly}, and explores integration of generative AI into art-based and storymaking modalities of expressive therapy~\cite{liu2024seahorse, du2024deepthink}.

We approach the reading that ``AI elicits a more authentic self'' with caution. Speech directed at an AI is itself a performance, shaped by a different set of social pressures than speech directed at another human---and the direction of those pressures varies across users and contexts~\cite{ho2018psychological, croes2024digital}. Rather than asking which context is more truthful, this paper attends to the systematic differences between the two as a potential source of diagnostic information.

We name this difference ``The Interaction Gap'' and propose it as a diagnostic \emph{lens} within drama therapy. Drama therapy places \emph{role} and \emph{aesthetic distance} at the center of its therapeutic mechanism~\cite{landy1996essential, jennings1998introduction}, with patients meeting themselves through a role rather than addressing their own material directly. Comparing self-presentation in front of an AI partner and a human partner extends this tradition: a second, comparable role context added to a practice that already works with role-mediated disclosure.

This paper contributes (1) a gap-as-lens framing that treats AI not as a co-creative material or a bridge to human care, but as a comparable interaction partner against which differences in self-presentation become legible; (2) task design considerations for eliciting the gap within drama therapy; (3) candidate measurement signals for the gap; and (4) two provocations for the workshop---the observer effect introduced by measurement, and the question of whether the lens is therapist-facing or patient-facing.

\section{Background}

\subsection{AI in Arts Therapy and Self-Disclosure}

Generative AI has been introduced into expressive arts therapy in several roles. As expressive material, image-based generative tools have been used in family storymaking sessions led by therapists~\cite{liu2024seahorse}, and human--AI co-creation has been probed within digital art therapy practice~\cite{du2024deepthink}. A common thread is that AI is treated as a \emph{material} or \emph{co-creator}, expanding what the patient can produce in session.

A parallel literature documents that conversational dynamics differ between AI and human interlocutors. Lucas et al.~\cite{lucas2014itsonly} showed that people disclose sensitive information more freely to a virtual interviewer than to a human one, an effect linked to perceived anonymity~\cite{croes2024digital}. The pattern extends beyond disclosure rates: comparisons of LLM chatbots and licensed therapists find that the two adopt systematically different communicative postures, with therapists eliciting more elaboration and chatbots leaning toward affirming and psychoeducational language~\cite{iftikhar2025comparison}. The picture is not uniform. Some studies find no significant difference in disclosure depth between chatbot and human partners under controlled conditions~\cite{ho2018psychological}, suggesting that the effect depends on agent embodiment, anonymity cues, and the user's expectations.

Building on this asymmetry, several recent systems position AI as a \emph{bridge} into human-led care. Park et al.~\cite{park2020mediator} designed a chatbot as a mediator that promotes deep self-disclosure before contact with a mental health professional. Quan et al.~\cite{quan2025relational} extend this framing for marginalized clients, treating LLM chatbots as \emph{boundary objects} that mediate knowledge gaps and power asymmetries between client and therapist. We explore a position adjacent to but distinct from the bridge framing. Rather than asking how AI smooths entry into human therapy, we ask whether the difference between AI and human interactions carries diagnostic information within drama therapy practice.

Within the intersection of AI and theater more broadly, recent work explores LLM-augmented improvisation and scriptwriting~\cite{kang2025tlp} and reflective tools in actor training~\cite{kang2026actorsnote}. These are not therapeutic settings, but they share with drama therapy a central reliance on \emph{role} as the medium of expression---an overlap we return to in §3.

\subsection{Drama Therapy}

The therapeutic mechanism of drama therapy is often described through the concept of \emph{aesthetic distance}~\cite{landy1983aesthetic, landy1996essential}. When a patient is too close to their own material---\emph{under-distanced}---they may be overwhelmed by affect; when too distant---\emph{over-distanced}---they may remain in intellectual analysis without emotional contact. Therapeutic insight is associated with the space between these, a position where the patient meets themselves through a role while simultaneously recognizing it as their own. By voicing material through a fictional character rather than directly, defenses soften and access widens. Our proposal builds on the possibility that this role-mediated mechanism continues to operate in a different form, when the audience for the role is an AI partner rather than a human one.

\section{Motivation}
This paper's motivation comes from our prior work, \emph{Actor's Note}~\cite{kang2026actorsnote}, which examined AI-generated questions used in character journaling during actor training. Across interviews, participants consistently described their interactions with the AI as carrying a different texture from interactions with human collaborators. P22 said, ``When analyzing scripts with people, it can feel like you must argue and prevail logically, which is exhausting. With AI, there's no need for that, which was helpful.'' P17 reported, ``When working with people, I sometimes couldn't fully express my thoughts.'' P1 described the AI as taking on an outside view: ``An actor's performance is ultimately for the audience, so it was helpful that the AI's questions felt like they came from an audience perspective.'' Read together, these accounts point to a tendency—rather than a uniform effect—for AI to defer some of the social friction characteristic of human-to-human interaction: the pressure to be logically dominant, to manage others' reactions, and to perform an idealized version of oneself. Actors are not patients. They engage with role-play professionally, with different stakes and different vulnerabilities from those entering a therapeutic setting. We treat the \emph{Actor's Note} finding as a \emph{generative observation} rather than as evidence transferable to clinical populations. Because drama therapy already uses \emph{role} as a therapeutic medium, the deferral of social friction reported by actors offers a point of entry---not an answer---for asking whether a comparable mechanism shapes patient self-presentation when an AI partner enters the room.

\section{The Interaction Gap as a Diagnostic Lens}
\subsection{The Gap as Systematic Divergence}

We define the Interaction Gap as the difference in a patient's self-presentation, observed consistently within that patient, when the same drama therapy task is performed with an AI partner and with a human partner. Self-presentation here includes verbal signals—word choice, depth of disclosure, hedging, topic avoidance—and non-verbal signals—hesitation, vocal prosody, gesture, and spatial use of the body. Both interactions are performances in the dramaturgical sense [4], each shaped by its own social pressures: the patient's reading of audience evaluation, the perceived stakes of the utterance, and the imagined relationship with the partner.

The Gap manifests differently across individuals—some patients may hedge more in front of an AI, others in front of a human—but is expected to be patterned within a given patient across sessions.

\subsection{Why Drama Therapy}

Drama therapy uses role-mediated disclosure as a core therapeutic mechanism~\cite{landy1996essential}. Introducing an AI partner adds a second, comparable role context to a practice that already works in this register. With the patient playing the same role in front of two different types of audience, the lens makes visible how the kind of audience shapes a role-mediated self-presentation.

Aesthetic distance contributes to this. When a patient voices the lines of a character rather than their own material, the defenses that would otherwise tighten around personal disclosure tend to loosen, and the gap becomes more legible. The patient holds the protective frame of ``this is the character's line, not mine'' while engaging with both audiences. The role-as-distance stays roughly constant across the two sessions, while the audience---the partner---varies. This is the comparison we are interested in.

We position this proposal as a \emph{lens} rather than a fixed framework. Developing it into a framework would require, at minimum, (a) clear operationalization of measurement units, (b) validation in clinical populations rather than adjacent ones, and (c) a defined mode of integration into therapeutic practice. These are themselves matters for the workshop to consider.

\section{Design Considerations}

\subsection{Task Design}

Eliciting the Interaction Gap in a drama therapy context requires tasks that meet several conditions.

(1) The patient performs from within the position of a fictional character rather than speaking directly about their own experience. For example: \emph{``You are Character A, who has been unjustly accused. Voice your defense.''} Grounded in Landy's distancing theory~\cite{landy1983aesthetic}, this lets the patient operate behind the protective frame of ``this is the character's line, not mine,'' under which the defenses that usually rise around direct disclosure tend to soften.

(2) The AI partner and the human partner engage the patient in the same role and scenario. Counterbalancing and temporal spacing between the two sessions are needed to control for order effects and learning effects. Minor variations of the same scenario (e.g., Character A $\rightarrow$ Character A$'$) can also be considered.

(3) Drama therapy is multi-modal---voice, body, space, expression. A text-only AI partner session collapses the comparable signal space sharply. For the gap to be measured beyond verbal signals, the AI partner should support voice and visual presence at some level of parity with the human session.

(4) Drama therapy ethics require that patients exit the role at the end of a session through a \emph{de-roling} protocol. The same standard applies to the AI partner session, and de-roling should be designed into the AI system rather than treated as a session-level afterthought.

\subsection{Measurement}

Candidate signals for the gap fall into three families.

(1) Lexical signals, captured through tools such as LIWC~\cite{pennebaker2015liwc}, include self-reference frequency, hedging markers (e.g., \emph{``I think,''} \emph{``maybe''}), affective vocabulary diversity, and social word use. These are well-established but limited to text-level analysis.

(2) Pragmatic signals---self-disclosure depth~\cite{altman1973social}, topic avoidance, and question--response asymmetry---capture semantic depth beyond word counts. They require qualitative coding, but they can surface structural differences in how the patient inhabits the role between the two audiences.

(3) Multi-modal signals align with the embodied character of drama therapy: vocal prosody (pitch, pace, pause length), gestural openness, and spatial proxemics (distance to and orientation toward the partner). Whether AI partner sessions can capture these signals at parity with human sessions is itself a design question worth raising.

The direction of divergence is read as informative regardless of which way it runs. Whether a patient hedges more in front of the AI or more in front of the human, the framing measures the difference between the two distributions.

\section{Discussion}
Two larger questions surround this proposal and remain open for workshop discussion.

\subsection{The Observer Effect and the Privacy Paradox}

Measuring the gap requires that the AI partner session be analyzable in some form. Once a patient understands that ``my conversation with the AI will ultimately be analyzed by my human therapist,'' the AI may stop functioning as a space of deferred social friction. The AI session can come to be read as the therapist's surveillance camera, and defenses that would otherwise loosen may tighten again, collapsing the very gap the system aims to measure.

Two design directions follow from this paradox. One is asymmetric data flow---an architecture in which raw transcripts are not shared with the therapist, and only gap-level indicators are computed inside the system and surfaced as visual mappings (akin to privacy-preserving aggregation, where only derived indices---not raw content---are surfaced to the therapist). The other is honesty about measurement---making the fact of measurement explicit to the patient from the outset, and treating the new performance that emerges under awareness of measurement as a legitimate object of analysis in its own right. Each direction trades measurement richness against privacy preservation, and this trade-off would be the central design tension as the lens develops.

\subsection{Whose Lens? Therapist-facing vs. Patient-facing Design}

Gap analysis can serve a therapist seeking deeper understanding of a patient, or it can serve the patient as a reflective mirror on their own variability across contexts. The two positionings call for substantially different systems---in data flow, visualization, consent structure, and timing.

A therapist-facing design may strengthen clinical insight while exposing the system more directly to observer-effect and surveillance concerns. A patient-facing design---in which the patient observes and interprets their own gap---sits closer to the self-knowledge orientation of expressive arts therapy and may ease surveillance concerns. It introduces its own design questions in turn: how the system should guide the patient's interpretation of their own gap, and how to design against the risk that patients draw mistaken self-diagnoses from their own data.

In either direction, the lens is intended for reflection within a therapeutic context, and we do not propose its use as a phenotyping instrument that reduces patients to diagnostic labels. 

\section{Conclusion}
This paper proposed the Interaction Gap as a diagnostic lens for drama therapy, treating the systematic difference between AI and human interactions as a source of information about the social pressures a patient navigates across contexts. The lens leaves task design, measurement, and ethical positioning open. We hope expressive arts therapists, HCI researchers, and AI ethics researchers can use the workshop to explore directions in which the lens could develop, and the limits it should respect.

%% References
\bibliographystyle{ACM-Reference-Format}
\bibliography{dis26-workshop}

@inproceedings{liu2024seahorse,
  title     = {``When He Feels Cold, He Goes to the Seahorse'': Blending Generative {AI} into Multimaterial Storymaking for Family Expressive Arts Therapy},
  author    = {Liu, Di and Zhou, Hanqing and An, Pengcheng},
  booktitle = {Proceedings of the 2024 CHI Conference on Human Factors in Computing Systems},
  series    = {CHI '24},
  year      = {2024},
  publisher = {ACM},
  doi       = {10.1145/3613904.3642852}
}

@article{du2024deepthink,
  title   = {{DeepThInk}: Designing and probing human-{AI} co-creation in digital art therapy},
  author  = {Du, Xuejun and An, Pengcheng and Leung, Justin and Li, April and Chapman, Linda E. and Zhao, Jian},
  journal = {International Journal of Human-Computer Studies},
  volume  = {181},
  pages   = {103139},
  year    = {2024},
  doi     = {10.1016/j.ijhcs.2023.103139}
}

@article{lucas2014itsonly,
  title   = {It's only a computer: Virtual humans increase willingness to disclose},
  author  = {Lucas, Gale M. and Gratch, Jonathan and King, Aisha and Morency, Louis-Philippe},
  journal = {Computers in Human Behavior},
  volume  = {37},
  pages   = {94--100},
  year    = {2014},
  doi     = {10.1016/j.chb.2014.04.043}
}

@article{ho2018psychological,
  title   = {Psychological, relational, and emotional effects of self-disclosure after conversations with a chatbot},
  author  = {Ho, Annabell and Hancock, Jeff and Miner, Adam S.},
  journal = {Journal of Communication},
  volume  = {68},
  number  = {4},
  pages   = {712--733},
  year    = {2018},
  doi     = {10.1093/joc/jqy026}
}

@article{croes2024digital,
  title   = {Digital Confessions: The Willingness to Disclose Intimate Information to a Chatbot and its Impact on Emotional Well-Being},
  author  = {Croes, Emmelyn A. J. and Antheunis, Marjolijn L.},
  journal = {Interacting with Computers},
  volume  = {36},
  number  = {5},
  pages   = {279--292},
  year    = {2024},
  doi     = {10.1093/iwc/iwae016}
}

@article{iftikhar2025comparison,
  title   = {A Comparison of Responses from Human Therapists and Large Language Model--Based Chatbots to Assess Therapeutic Communication: Mixed Methods Study},
  author  = {Iftikhar, Zainab and Ramos, Sean and Suh, Eunseo and Schiavi, Lisa and Mishra, Aishik and Lawrence, Maya and Welton, Emory and Yang, Mendy and Park, Pia Sukyung and Rivero, Jamilex and Huang, Ryan and Yang, Lillian and Bickmore, Timothy and Huang, Jeff},
  journal = {JMIR Mental Health},
  volume  = {12},
  pages   = {e69709},
  year    = {2025},
  doi     = {10.2196/69709}
}

@article{siddals2024happened,
  title   = {``It happened to be the perfect thing'': experiences of generative {AI} chatbots for mental health},
  author  = {Siddals, Steven and Torous, John and Coxon, Astrid},
  journal = {npj Mental Health Research},
  volume  = {3},
  number  = {1},
  pages   = {48},
  year    = {2024},
  doi     = {10.1038/s44184-024-00097-4}
}

@article{park2020mediator,
  title   = {Designing a Chatbot as a Mediator for Promoting Deep Self-Disclosure to a Real Mental Health Professional},
  author  = {Park, SoHyun and Choi, Jeewon and Lee, Sungwoo and Oh, Changhoon and Kim, Changdai and La, Soohyun and Lee, Joonhwan and Suh, Bongwon},
  journal = {Proceedings of the ACM on Human-Computer Interaction},
  volume  = {4},
  number  = {CSCW1},
  articleno = {31},
  year    = {2020},
  doi     = {10.1145/3392836}
}

@article{quan2025relational,
  title         = {Relational Mediators: {LLM} Chatbots as Boundary Objects in Psychotherapy},
  author        = {Quan, Jiatao and Li, Ziyue and Zhu, Tian Qi and Li, Yuxuan and Wang, Baoying and Pratt, Wanda and Gao, Nan},
  journal       = {arXiv preprint arXiv:2512.22462},
  year          = {2025},
  eprint        = {2512.22462},
  archivePrefix = {arXiv}
}

@inproceedings{kang2026actorsnote,
  title     = {Actor's Note: Examining the Role of {AI}-Generated Questions in Character Journaling for Actor Training},
  author    = {Kang, Sora and Zoh, Jaemin and Kim, Hyoju and Park, Hyeonseo and Lim, Hajin and Lee, Joonhwan},
  booktitle = {Proceedings of the 2026 CHI Conference on Human Factors in Computing Systems},
  series    = {CHI '26},
  articleno = {1286},
  numpages  = {19},
  year      = {2026},
  publisher = {Association for Computing Machinery},
  address   = {New York, NY, USA},
  isbn      = {9798400722783},
  doi       = {10.1145/3772318.3790370},
  url       = {https://doi.org/10.1145/3772318.3790370}
}

@inproceedings{kang2025tlp,
  title     = {Theatrical Language Processing: Exploring {AI}-Augmented Improvisational Acting and Scriptwriting with {LLMs}},
  author    = {Kang, Sora and Lee, Joonhwan},
  booktitle = {Proceedings of the International Symposium on Electronic/Emerging Art (ISEA 2025)},
  pages     = {759--766},
  year      = {2025},
  month     = {7},
  address   = {Seoul, Republic of Korea},
  publisher = {Art Center Nabi, Institute for Culture and Arts Seoul National University},
  doi       = {10.23362/KOEN2025.07.25.1.099}
}

@article{landy1983aesthetic,
  title   = {The use of distancing in drama therapy},
  author  = {Landy, Robert J.},
  journal = {The Arts in Psychotherapy},
  volume  = {10},
  number  = {3},
  pages   = {175--185},
  year    = {1983},
  doi     = {10.1016/0197-4556(83)90006-0}
}

@book{landy1996essential,
  title     = {Essential Drama Therapy: Theories, Methods, and Practices},
  author    = {Landy, Robert J.},
  year      = {1996},
  publisher = {Jessica Kingsley Publishers},
  address   = {London}
}

@book{altman1973social,
  title     = {Social Penetration: The Development of Interpersonal Relationships},
  author    = {Altman, Irwin and Taylor, Dalmas A.},
  year      = {1973},
  publisher = {Holt, Rinehart and Winston},
  address   = {New York}
}

@techreport{pennebaker2015liwc,
  title       = {The Development and Psychometric Properties of {LIWC2015}},
  author      = {Pennebaker, James W. and Boyd, Ryan L. and Jordan, Kayla and Blackburn, Kate},
  institution = {University of Texas at Austin},
  year        = {2015},
  address     = {Austin, TX}
}

@book{malchiodi2005expressive,
  title     = {Expressive Therapies},
  editor    = {Malchiodi, Cathy A.},
  year      = {2005},
  publisher = {Guilford Press},
  address   = {New York}
}

@book{jones2007drama,
  title     = {Drama as Therapy: Theory, Practice and Research},
  author    = {Jones, Phil},
  year      = {2007},
  edition   = {2nd},
  publisher = {Routledge},
  address   = {London}
}

@book{jennings1998introduction,
  title     = {Introduction to Dramatherapy: Theatre and Healing -- Ariadne's Ball of Thread},
  author    = {Jennings, Sue},
  year      = {1998},
  publisher = {Jessica Kingsley Publishers},
  address   = {London}
}

\end{document}